\documentclass{osa-article}

\journal{oe}

\articletype{Research Article}

\usepackage{lineno}

\begin{document}

\title{Single-shot quantitative aberration and scattering length measurements in mouse brain tissues using an extended-source Shack-Hartmann wavefront sensor}

\author{Sophia Imperato,\authormark{1,2,*} Fabrice Harms,\authormark{3} Antoine Hubert,\authormark{1,3} Mathias Mercier, \authormark{1} Laurent Bourdieu,\authormark{2} and Alexandra Fragola\authormark{1}}

\address{\authormark{1}Laboratoire de Physique et Etude des Matériaux, ESPCI Paris, Université PSL, CNRS, Sorbonne Université, 
10 rue Vauquelin, 75005 Paris, France\\
\authormark{2}Institut de Biologie de l’ENS (IBENS), Département de biologie, École normale supérieure, CNRS, INSERM, Université PSL, 
46 rue d’Ulm, 75005 Paris, France\\
\authormark{3}Imagine Optic, 18 rue Charles de Gaulle, 91400 Orsay, France}

\email{\authormark{*}sophia.imperato@espci.fr} 



\begin{abstract}
Deep fluorescence imaging in mammalian brain tissues remains challenging due to scattering and optical aberration-induced loss in signal and resolution. Correction of aberrations using adaptive optics (AO) requires their reliable measurement in the tissues. Here, we show that an extended-source Shack-Hartmann wavefront sensor (ESSH) allows quantitative aberration measurements through fixed brain slices with a thickness up to four times their scattering length. We demonstrate in particular that this wavefront measurement method based on image correlation is more robust to scattering compared to the standard centroid-based approach. Finally, we obtain a measurement of the tissue scattering length taking advantage of the geometry of a Shack-Hartmann sensor. 
\end{abstract}

\section{Introduction}
Optical microscopy has emerged as a major tool in neuroscience as it allows to image the architecture of neuronal networks and to record their functional activity in several animal models as zebrafish, rodents, ferrets or primates through different methods including widefield fluorescence microscopy \cite{Gilad_Gallero-Salas_Groos_Helmchen_2018}, confocal \cite{Dussaux_Szabo_Chastagnier_Fodor_Leger_Bourdieu_Perroy_Ventalon_2018}, one-photon \cite{Ahrens_Orger_Robson_Li_Keller_2013, Hillman_Voleti_Li_Yu_2019}  or two-photon light-sheet microscopy \cite{Wolf_Supatto_Debregeas_Mahou_Kruglik_Sintes_Beaurepaire_Candelier_2015} and non-linear scanning microscopy \cite{Helmchen_Denk_2005,Lecoq_Orlova_Grewe_2019, Yang_Yuste_2017}. Modern optical sectioning methods such as confocal or light-sheet can image a large part of the entire brain of relatively transparent animal models such as zebrafish, drosophila or C-Elegans. In mammalian brains, depth penetration is limited even at low depth by scattering and imaging can only be achieved using non-linear microscopy. However, at large depths, optical aberrations linked to the refractive index inhomogeneity of biological tissues still limit resolution and signal intensity in all microscopy modalities\cite{Booth_2014,Ji_2014}. To overcome this difficulty, adaptive optics (AO) has been implemented on several linear and non-linear optical sectioning microscopy setups and currently provides a reliable live correction of the aberrations\cite{Rodriguez_Ji_2018}, enabling functional imaging of synaptic boutons\cite{Sun_Tan_Mensh_Ji_2016}, axons and spines\cite{Liu_Li_Marvin_Kleinfeld_2019} and soma\cite{Wang_Sun_Richie_Harvey_Betzig_Ji_2015} in infragranular layers of the mouse cortex in 2-photon microscopy, or enabling subcellular imaging of organelle dynamics in the early zebrafish brain in light-sheet microscopy\cite{Liu_Upadhyayula_Milkie_Singh_Wang_Swinburne_Mosaliganti_Collins_Hiscock_Shea_2018}, and deeper imaging of multicellular tumor spheroids\cite{Jorand_Le_Corre_Andilla_Maandhui_Frongia_Lobjois_Ducommun_Lorenzo_2012,Masson_Escande_Frongia_Clouvel_Ducommun_Lorenzo_2015}. 
The first strategy to implement AO in microscopy is based on a sensorless configuration\cite{Debarre_Botcherby_Watanabe_Srinivas_Booth_Wilson_2009,Facomprez_Beaurepaire_Debarre_2012,Wang_Liu_Milkie_Sun_Tan_Kerlin_Chen_Kim_Ji_2014,Rodriguez_Chen_Rivera_Mohr_Liang_Natan_Sun_Milkie_Bifano_Chen_2021}, that does not make use of a wavefront sensor but analyses the variation of the fluorescence signal induced by a wavefront modulator to drive the correction of aberrations. This method minimizes instrumental complexity, and provides a good resilience to scattering. However, it relies on a time-consuming iterative approach requiring many acquisitions before reaching a good correction: typically, 30 seconds are necessary for a single iteration of a sensorless process, and a couple of iterations are required to reach an optimal correction. The approach is thus hardly compatible with time-varying aberrations and with photobleaching issues, in particular when multiple corrections in a volume are required, e.g. for a small isoplanetic patch which is often the case in depth\cite{Liu_Upadhyayula_Milkie_Singh_Wang_Swinburne_Mosaliganti_Collins_Hiscock_Shea_2018}. 
The second approach to enable AO in microscopy is based on direct wavefront sensing, and makes use of a wavefront sensor to allow a fast and accurate convergence of the AO loop, with a better photon budget: direct wavefront sensing is thus a key method to evaluate and correct aberrations \emph{in vivo} over large scales. Direct wavefront sensing for AO-enhanced optical microscopy has been reported over the last years through several approaches, using either Shack-Hartmann (SH) sensors\cite{Liu_Li_Marvin_Kleinfeld_2019,Wang_Sun_Richie_Harvey_Betzig_Ji_2015,Tao_Azucena_Fu_Zuo_Chen_Kubby_2011,Hubert_Harms_Juvenal_Treimany_Levecq_Loriette_Farkouh_Rouyer_Fragola_2019,Jorand_Le_Corre_Andilla_Maandhui_Frongia_Lobjois_Ducommun_Lorenzo_2012,Liu_Upadhyayula_Milkie_Singh_Wang_Swinburne_Mosaliganti_Collins_Hiscock_Shea_2018}, or partitioned aperture wavefront (PAW) sensors\cite{Li_Beaulieu_Paudel_Barankov_Bifano_Mertz_2015} as a variant of pyramid wavefront sensors applicable to microscopy. The use of PAW in fluorescence microscopy is severely limited, since it requires only a moderate spatial incoherence of the excitation source, so that its demonstration in fluorescence microscopy was restricted to widefield using a specific illumination geometry (Oblique Back Illumination) \cite{Li_Beaulieu_Paudel_Barankov_Bifano_Mertz_2015,Li_Bifano_Mertz_2016}. Direct wavefront sensing in microscopy is thus currently mostly based on SH sensors. Early AO demonstrations based on SH\cite{Tao_Azucena_Fu_Zuo_Chen_Kubby_2011,Jorand_Le_Corre_Andilla_Maandhui_Frongia_Lobjois_Ducommun_Lorenzo_2012} used extrinsic fluorescent beads injected into the sample to provide a guide star for centroid computation, which is usually unwanted. More recently, Wang et al. used two-photon excitation, scanned over a given field of view, to create a guide star outside the sample by descanning the fluorescence signal, the guide star being then used by a SH sensor based on a conventional centroid computation\cite{Wang_Sun_Richie_Harvey_Betzig_Ji_2015}. The approach was successfully used to drive AO in Lattice Light-Sheet \cite{Liu_Upadhyayula_Milkie_Singh_Wang_Swinburne_Mosaliganti_Collins_Hiscock_Shea_2018}, as well as in two-photon\cite{Wang_Sun_Richie_Harvey_Betzig_Ji_2015} and in Structured Illumination Microscopy (SIM) \cite{Li_Zhang_Chou_Newman_Turcotte_Natan_Dai_Isacoff_Ji_2020}. Recently, we demonstrated the use of an Extended-Source SH wavefront sensor (ESSH) to enable AO in light-sheet microscopy\cite{Hubert_Harms_Juvenal_Treimany_Levecq_Loriette_Farkouh_Rouyer_Fragola_2019}, by adapting pioneer work from astronomy to the constraints of fluorescence microscopy. The method relies on the cross-correlation of images of an extended source obtained through a microlens array. In fluorescence microscopy, when coupled to an optical sectioning method such as two-photon microscopy or light-sheet fluorescence microscopy, the ESSH uses the fluorescence signal as a guide plane, providing lower instrumental complexity and cost than the previous approach based on the scan and descan of a non-linear signal. Its efficiency has been proven when coupled to light-sheet for AO-enhanced neuroimaging in the adult drosophila brain in weekly scattering conditions\cite{Hubert_Harms_Juvenal_Treimany_Levecq_Loriette_Farkouh_Rouyer_Fragola_2019}. 

However, direct wavefront sensing in scattering tissue is limited in depth penetration. Indeed, in pioneering SH wavefront measurements based on centroid estimation in the rodent brain, the strong scattering of the emitted fluorescence reduces significantly the capacity to measure the centroid position due do the decreasing ballistic signal and increasing background on images below each lenslet \cite{Wang_Sun_Richie_Harvey_Betzig_Ji_2015,Liu_Li_Marvin_Kleinfeld_2019}. A method for direct wavefront sensing more resilient to scattering of the fluorescence emission would therefore improve the use of AO in optical microscopy. To this aim, we analyse here the performances of an ESSH in scattering tissues and show that it provides quantitative aberration measurements in highly scattering fixed mouse brain tissues, together with a fast and precise estimation of its scattering length. We compare our extended-source wavefront measurement approach to the Shack-Hartmann method based on centroid calculation, and show the benefit of the former in the case of scattering samples, and its improved accuracy at large depths in low Signal to Background Ratio conditions. In a proof of principle experiment, the ESSH method is finally used with a model sample consisting of neurons in culture placed under a fixed brain slice to illustrate the image quality improvement that AO, driven by ESSH, can provide in microscopy experiments in brain tissue at large depths, and in particular in the mouse cortex.

\section{Material and methods}\label{Methods}

\subsection{Set-up}
\begin{figure}[ht]
    \centering
    \includegraphics[width=13cm]{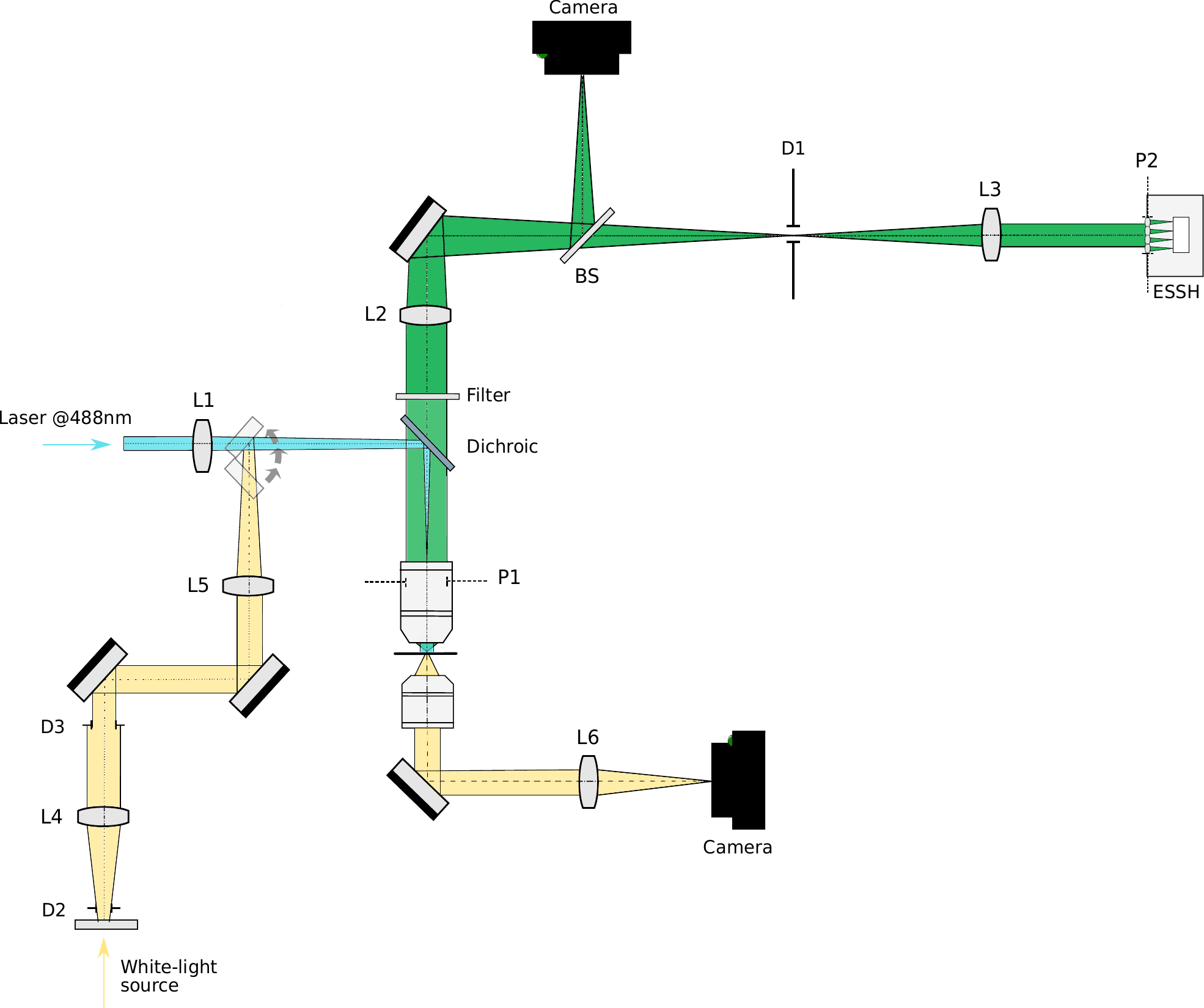}
    \caption{Schematic of the set up. L1-6 lenses, D1 squared field diaphragm, D2 aperture diaphragm, D3 field diaphragm, BS 50:50 beamsplitter, P1 back focal plane of the objective, conjugated to P2 located on the microlens array of the ESSH.}
    \label{fig:Fig1}
\end{figure}

An ESSH wavefront sensor was implemented on a custom-made epifluorescence set-up represented in Fig.\ref{fig:Fig1}, composed of a laser light source at 488 nm (Cobolt 06-MLD) focused in the back focal plane of the water immersion objective (Olympus 20X, NA=1, XLUMPFLN20XW). The fluorescence emitted was collected through the objective, its back focal plane being conjugated with the ESSH microlens array by a pair of achromatic lenses (focal lengths: 300 mm and 80 mm). The ESSH was similar to the device reported in our previous publication\cite{Hubert_Harms_Juvenal_Treimany_Levecq_Loriette_Farkouh_Rouyer_Fragola_2019}, and was composed of an array of 17 x 23 microlenses, with a focal length of 5.1 mm, the CMOS sensor being in the focal plane of the microlenses. A squared field diaphragm was placed in front of the ESSH, leading to a field of view of 120 x 120 µm² in the object plane, for each microlens, thus selecting the correlation area and avoiding crosstalk between adjacent thumbnails. A 50:50 beamsplitter (BS) separated the beam between a full-field camera (Thorlabs) and the ESSH path. The observation wavelength was selected by either two bandpass filters (525/50, 625/40 from Chroma) or a long-pass filter (LP715 from Semrock) allowing us to measure the wavefront at three different wavelengths using three types of emitters (see details in ‘Sample preparation’ section) being deposited on a single coverslip. A white-light transmission imaging path was added in order to check sample positioning, in particular to localize precisely the cortex area of the coronal slices for wavefront measurement.

The set-up used for closed-loop AO, corresponding to the results presented on Fig.4, was composed of a water immersion objective (Leica 25X, NA=0.95), two pair of achromatic relay lenses to conjugate the back focal plane of the objective on a Mirao52e deformable mirror (Imagine Eyes), and then on the ESSH. The scientific image was realised on an ORCA Flash V2 (Hamamatsu). The correlation computation was performed over the 130 x 130 µm² central zone of the full-field image.

The sample was imaged by the ESSH thumbnails onto the camera sensor, leading to undersampled thumbnails. A reference thumbnail was chosen centrally, and correlations were computed between each thumbnail and the reference one in order to retrieve the relative displacement of each thumbnail to the reference. A pre-processing step was required to avoid correlation errors induced by noise. Considering the low SBR on our thumbnails (see SBR definition below), a Laplacian of Gaussian filter was selected. A pyramidal fit was used to find the correlation peak with sub-pixelic displacement for each thumbnail, leading to the local displacement of the wavefront; this choice results in a trade-off between accuracy and time calculation. These steps followed the procedure and pseudo-code described in Anugu et al\cite{Anugu_Garcia_Correia_2018}. Finally, the wavefront reconstruction was achieved with a conventional zonal method.
The closed-loop was run at a 10Hz rate, with the following computer configuration: processor IntelCore i5-8400 CPU@2.8GHz, 32Go RAM, Windows 64bits.

\subsection{Sample preparation}
Because of the low numerical aperture (NA) of the microlenses (NA = 0.03), the ESSH required to be coupled with an optical sectioning method to provide quantitative wavefront measurements in thick 3D samples. To demonstrate the performances of the ESSH in the case of a scattering sample, we used a simple configuration, where the sectioning was inherent to the samples: we used 2D fluorescent samples placed below unlabelled 3D scattering samples (fixed mouse brain slices) and achieved by this mean wavefront measurement on an epi-fluorescence microscope lacking optical sectioning. This approach allowed to easily control the scattering properties of the sample using slices of known thicknesses, which would be extremely challenging using intact biological samples, and relaxed the need for implementing optical sectioning in the instrument. Such a model has already been used in other AO-based experiments \cite{Siemons_Hanemaaijer_Kole_Kapitein_2021}.

Experimental procedures were conducted in accordance with the institutional guidelines and in compliance with French and European laws and policies and with the ARRIVE guidelines\cite{Sert_Hurst_Ahluwalia_Alam_Avey_Baker_Browne_Clark_Cuthill_Dirnagl_2020}. All procedures were approved by the ‘Charles Darwin’ local institutional ethical committee registered at the French National Committee of Ethical Reflection on Animal Experimentation under the number 05 (authorization number: APAFIS 26667).

Three 6-months old C57BL6 male mice were sacrificed by an overdose of Euthasol, after being placed under deep sedation by overdose of isoflurane (5 min at 5\% isoflurane in an induction box). The extracted brain was then stored overnight in a solution of 4\% paraformaldehyde and finally rinsed in phosphate buffer solution (PBS). Coronal slices of different thicknesses ranging from 50 µm to 300 µm were then cut and stored in PBS.

Several fluorescent emitters were used; 2-µm fluorescent beads (ThermoFischer, $\lambda_{em}$ = 515 nm) were selected as emitters within the green range while quantum dots (QD) aggregates were used as emitters in the red and near-infrared ranges in order to keep the same excitation wavelength. ZnCuInSe/ZnS QD emitting at 800 nm were synthesized following the protocol reported by Pons et al\cite{Pons_Bouccara_Loriette_Lequeux_Pezet_Fragola_2019} whereas those emitting at 614 nm were obtained from protocol described by Yang et al\cite{Yang_Wu_Williams_Cao_2005}. QD in hexane (4 nmol) were mixed with ethanol (1 mL) to precipitate QD by centrifugation. QD were resuspended in chloroform (250 µL) and CTAB (5 mL, 3 mM) was added. The solution was heated for 20 min at 80°C to evaporate chloroform. The QD aggregates surrounded by hydrophobic ligands were precipitated by centrifugation (5 min, 8500 rpm) and then dispersed into ethanol. A second round of centrifugation was necessary to eliminate all non-aggregated QD. Finally, aggregates were dispersed into ethanol (1 mL). The aggregate solutions were diluted 10 times. 500-nm aggregates were obtained with respectively an emission peak at 614 nm (FWHM = 32 nm) for the visible ones, and 805 nm (FWHM = 17 nm) for the near-IR ones.

Samples were prepared with the three types of emitters. 4 µL of the 2-µm commercial beads, diluted 1/5000e, were deposited first on a 170-µm-thick coverslip then dried up for 5 min at 60°C. The two types of aggregates were deposited on the same side of the coverslip (4 µL of each). On the other side of the coverslip a fixed brain slice was placed, with spacers. Another coverslip was added on top to avoid the adhesion of the brain slice to the water-immersion objective during the experiment.

Neuronal cultures containing a mVenus fluorescent tag were generated by an ex-utero electroporation in the sensory cortex of E15 murin embryos. A drop of DNA pH1SCV2\_shRNA\_control (0.7 µg/µl) was injected with a capillary in one hemisphere of the sensory cortex of the embryo and then electroporated. This region was then extracted from the whole brain and dissected in a medium containing 10\% HBSS, 5\% glucose and 2\% HEPES. The piece of remaining cortical hemisphere was digested by chemical and mechanical digestions, first, under the action of papain for 15 min at 37°C and then under pipet up and down movements, to obtain individual neurons. The neurons were plated onto a 18-mm coverslip coated with 80 µg/ml polyornithine in 12 well plates, at a density of 180,000 cells/well in MEM, 10\% HS, 1\% glutamax and 1\% sodium pyruvate 100 mM and incubated for 2 hours at 37°C, 5\% CO2. When the neurons were completely adherent on the coverslip, the medium was replaced by Neurobasal, 2\% B27, 1\% glutamax and antibiotics. One third of the medium was changed every 5-6 days.

To obtain model scattering samples, fluorescent emitters (2-µm commercial beads diluted 1/5000e (4 µL), dried up for 5 min at 60°C) were first deposited on a coverslip. On the other side, 2-µm polystyrene beads (Sigma-Aldrich, 78452) dispersed in water were used as scattering samples and deposited. 1 µL of the commercial solution, containing 10\% of beads in mass was diluted 10 times to obtain a solution of 0.002 bead/µm$^{3}$ with a scattering length of 74.9 µm. Similarly, a 5-times diluted solution (0.005 bead/µm$^{3}$) with a scattering length of 29.9 µm was realised. The scattering coefficients were calculated with Mie-scattering theory, at 515 nm (online calculator \url{https://omlc.org/calc/mie_calc.html}). 10 µL of the diluted solutions were deposited on the coverslip, surrounded by 130-µm-thick spacers, and recovered by another coverslip to retain the scattering droplet.

\subsection{Third order Spherical Aberration (SA3) measurement and modelling}
SA3 coefficient is described in the Zernike polynomials as $6\rho^{4}-6\rho^{2}+1$. Thus, the focus variation has a strong impact on the spherical aberration value, as a best focus can be found to minimize the later. In order to eliminate this factor, we minimized experimentally the focus aberration to eliminate this bias from our analysis. Only the contribution of the brain slice was considered here: a reference wavefront, corresponding to the aberrations induced by the instrument and the coverslips, was subtracted at each wavelength. For each brain slice thickness, at least seven acquisitions were realised in different areas of the cortex, at the three considered wavelengths simultaneously.

We simulated the optical system in Zemax in order to obtain an order of magnitude of the refractive index mismatch–induced spherical aberration for mouse brain slices. A simple model consisting in two additional interfaces inserted between the emitters’ plane and the objective was chosen. The brain slice was modelled by a layer of refractive index of 1.368, previously experimentally measured on fixed mouse brain slices and reported by Lue et al\cite{Lue_Bewersdorf_Lessard_Badizadegan_Dasari_Feld_Popescu_2007}. The process was repeated for the corresponding experimental thicknesses used. 

\subsection{Numerical simulation of the measured shift of a thumbnail for several SBR}
The SBR of a given thumbnail was defined as the ratio between the average signal value of the 1\% brightest pixels, and the background mean value, considered as the mean value of the 80\% dimmest pixels. 
A well-sampled point spread function corresponding to a squared aperture, at 600 nm, was first generated numerically. Poissonian noise was added with a variance equal to the square root of the signal plus the background value. The generated image was then undersampled on one hand to obtain our reference thumbnail, with dimensions matching to the experimental ones. On the other hand, the equivalent of a 0.5 pixel shift was introduced in the two directions and the shifted image was undersampled afterwards. As discussed in Anugu et al\cite{Anugu_Garcia_Correia_2018} it corresponds to the minimum bias error value in the shift retrieval, for the centroid process as well as for the correlation-based method. The induced shift was then retrieved by the two methods computed here. The centroid position was measured on the two thumbnails, on which a threshold corresponding to the background value plus 10\% of the difference between the maximum and the background value was applied. The threshold choice was motivated by the results shown on Fig.4c and Fig.4d of the influence of the threshold choice on the shift retrieval in low SBR conditions. This simulation was run on fifty images for each SBR value. 
For the subpixelic shift retrieval on the experimental thumbnails, 35 x 35 pixels² images were first oversampled by a factor 100 and then shifted by the equivalent of 0.5 pixel in the two directions. After under-sampling the shifted images, the shift retrieval between the initial thumbnails and the shifted ones was performed as described above.

\section{Results}

\subsection{Signal to Background Ratio (SBR) across the ESSH wavefront sensor in scattering tissues.}
The ESSH microlenses are conjugated with the back focal plane of the objective (Fig.2a). Due to the scattering of light in biological samples, the images below each micro-lens are losing contrast (Fig.2b) in depth, as in the case of the centroid-based measurement \cite{Wang_Sun_Richie_Harvey_Betzig_Ji_2015,Liu_Li_Marvin_Kleinfeld_2019}. This loss in contrast ultimately limits the possibility to reliably measure a wavefront with Shack-Hartmann sensors at large depth in scattering tissues. More specifically, the intensity distribution on each thumbnail of the ESSH shows a decrease of the signal from the central thumbnail to the ones on the edges, with a rotational symmetry (Fig.2b). At the same time, the background decreases, but slower, from the centre to the edge of the ESSH (Fig.2b). Accordingly, the signal to background ratio (SBR) decreases significantly from the centre to the edge with the same rotational symmetry (Fig. 2b). On the contrary, in the absence of a scattering tissue and with a homogeneous fluorescent sample, the signal on the ESSH is uniform in intensity over all the thumbnails, except the ones on the most external row where the entrance of light rays starts being limited by the edge of the pupil (data not shown). This effect can be explained by the geometry of the sensor (Fig.2a): when a thick scattering sample is positioned on top of the imaged fluorescence emitters, the emitted signal received by the outer microlenses is propagating through a larger distance in the scattering sample than the one received by central microlenses. Let us note that this effect is not specific to the ESSH but is present in SH with centroid-based estimation of shifts too\cite{Wang_Sun_Richie_Harvey_Betzig_Ji_2015}. When measuring aberrations through scattering media, this Signal to Background (SBR) deterioration across the sensor limits thus the size of the pupil over which the wavefront can be computed, with the limiting SBR found at the outer thumbnails.

\subsection{Scattering coefficient measurements} 
Thanks to this intensity profile across the ESSH, the scattering length of the medium can be measured. In a scattering sample, the intensity of the ballistic fluorescence signal decays with depth following 
\begin{equation}
    I(z) \propto \exp{\left(-\frac{z}{l_{s}}\right)}
\end{equation}  where $l_{s}$ is the scattering mean free path (or scattering length) of the sample and $z$ its thickness. In the geometry of the ESSH, the thickness $z_{p}$ crossed to reach the thumbnail of index $p$ ($p$ varies between 0 (central thumbnail) and 7 (most external row of thumbnails) increases as 
\begin{equation}
    z_{p}=\frac{z_{c}}{\cos{\theta_{p}}}
\end{equation}
where $z_{c}$ is the slice thickness and $\theta_{p}$  is the angle over which the thumbnail $p$ is seen, with \begin{equation}
    \theta_{p} = \arctan{\left(\frac{p}{7}\tan{\theta_{max}}\right)}
\end{equation}
Measurements on a calibrated scattering sample composed of 2-µm polystyrene beads and 130-µm-thick are first performed to confirm the accuracy of the method, at two bead concentrations chosen to be within the range of $l_{s}$ values of fixed brain tissues. In these experiments, we position a single fluorescence emitter at 515 nm in the ESSH field of view. In Fig.2c, the intensity of the ballistic signal is plotted as a function of the effective thickness after the subtraction of the background on each thumbnail. From the slope of the semi-log plot, which is equal to $1/l_{s}$, we measure for a 0.005 sphere/µL concentration a scattering length of $26.5\pm 0.3$ µm, very close to the expected value of 29.9 µm. Similarly, we measure a $69.5\pm 1.4$ µm scattering length at 0.002 sphere/µL concentration where 74.9 µm was expected. This demonstrates that the scattering length can be measured using the ESSH sensor with an accuracy of 10\%.
Measurements on fixed mouse brain coronal slices, in the cortex, are performed at three different wavelengths (515 nm, 615 nm and 805 nm) to determine the chromatic dependence of the scattering length of the tissue. To improve the measurement accuracy, three slice thicknesses are used at each wavelength: at 515 nm and 615 nm, 50-, 100- and 150-µm-thick slices and at 805 nm, 100-, 150- and 200-µm-thick slices. The selected slice thicknesses correspond to ranges of $l_{s}$ to $2.5 l_{s}$. For each slice, the intensity is normalized to plot and fit the data for all slices on a single figure (Fig.2d); the ordinate of the individual semi-log plot being subtracted to each acquisition. The measured scattering coefficients are $38 \pm 2$ µm at 515 nm, $47 \pm 2$ µm at 615 nm and $77\pm 5$ µm at 805 nm, increasing significantly with the wavelength as expected. The obtained $l_{s}$ values at the three wavelengths are compared to previously reported scattering length values in the Discussion.

\begin{figure}[ht]
    \centering
    \includegraphics[width= 13cm]{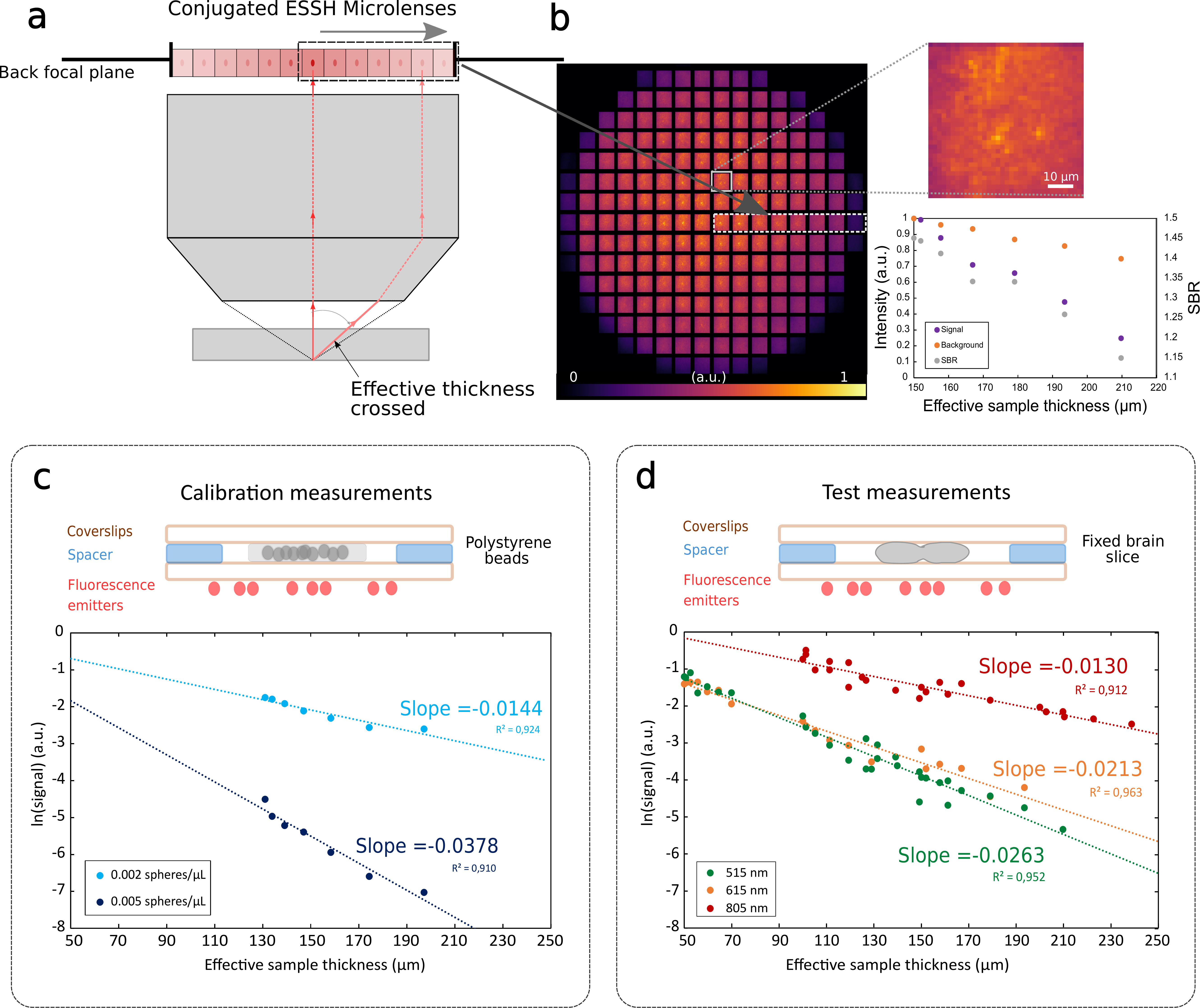}
    \caption{Scattering length measurement on fixed brain slices. (a) The ESSH microlenses are conjugated with the back focal plane of the objective. Each microlens sees the same field of view but with a different angle. (b) Example of an ESSH image showing the decrease in intensity between the side-by-side thumbnails within the doted boxes (represented in (a)), with a zoom on one corresponding thumbnail, obtained through a 150-µm-thick brain slice at 515 nm. Corresponding plot of the signal, the background and the signal to background as a function of the effective thickness seen along the direction of the doted boxes (c) Calibration measurements with the corresponding sample scheme. (d) Measurements on brain slices, with sample scheme.}
    \label{fig:fig2}
\end{figure}

\subsection{Spherical aberration measurements with ESSH.}\label{SA3}
In this section, we show that the ESSH sensor provides a reliable wavefront measurement in very scattering conditions. We consider in particular third-order Spherical Aberration (SA3), which is mainly generated by refractive index mismatch and which depends on the sample thickness, a parameter that we can easily adjust. We use 50- to 150-µm-thick slices within the visible range and up to 250-µm-thick in the near-IR. These thicknesses correspond to $1.3 l_{s}$ to $3.9 l_{s}$  at 515 nm, to $1.0 l_{s}$ to $3.3 l_{s}$ at 615 nm and to $0.6 l_{s}$ to $3.2 l_{s}$ at 805 nm. The increase in thickness corresponds to an increase in Optical Path Difference (OPD), that is the difference of refractive index between the tissue and water times the sample thickness. The variation of the tissue refractive index between the different wavelengths is neglected.
Comparisons of the experimental SA3 values to the one obtained by simulations (see Methods), presented in Fig.3a-c, show that our ESSH sensor is providing accurate aberration measurement and confirm quantitatively the increase of the spherical aberration with the brain slice thickness, in agreement with the simulation. At low slice thickness (50 µm), the SA3 induced by the sample is within the measurement error values of the ESSH (8-9 nm is predicted by the Zemax simulation). Our measurement method reaches its limits when the slice thickness became greater than $3.9 l_{s}$ with commercial beads as fluorescent emitters, and $3.3 l_{s}$ with the aggregates. Corresponding images on thumbnails are shown on Fig.3d. These slice thicknesses correspond to effective thicknesses crossed on the outer microlenses of $4.8   l_{s}$ and $5.8 l_{s}$, for which the ballistic signal is considerably reduced at the pupil edges (typ. 0.25\%), corresponding to a SBR of 1.3 only. Alongside this loss of ballistic signal, we observe a more important variability of the SA3 values for a single slice, as a consequence of the loss of slope accuracy evaluation at these SBR. Measuring aberrations with thicker slices could be achieved by restricting the pupil over which the wavefront is calculated, but these results could not be compared to the previous ones with thinner slices, performed on the entire pupil. Our ESSH thus allows performing quantitative aberration measurement through scattering slices that are up to $3.9 l_{s}$-thick. 

\begin{figure}[ht]
    \centering
    \includegraphics[width=13cm]{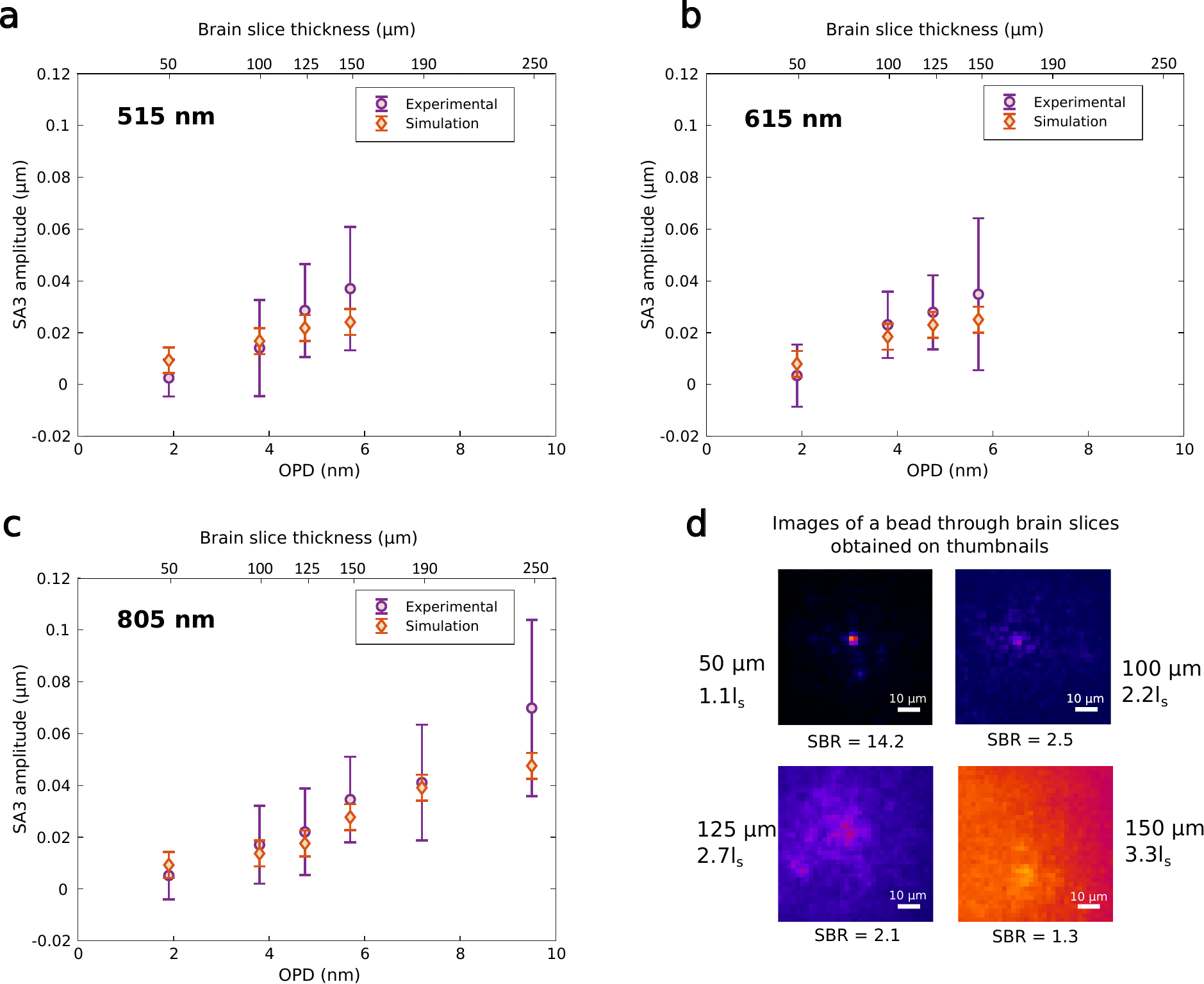}
    \caption{Spherical Aberration measurement on scattering sample with ESSH. (a)-(c) 3rd order Spherical Aberration amplitude increases with the Optical Path Difference (i.e. the brain slice thickness) induced by mouse brain slice at 3 wavelengths, mean and standard deviation plotted. The experimental data are compared to simulation data (described in Methods). (d) Examples of signal on external thumbnails for 4 different thicknesses of brain slice, at 615 nm. The corresponding values of slice thicknesses are given in terms of number of scattering lengths.}
    \label{fig:Fig3}
\end{figure}

\subsection{Comparison of ESSH/centroid computation at low SBR}\label{Comp_low_SBR}
To confirm that ESSH measurements are resilient to strong scattering conditions, in particular in comparison to the centroid-based measurements, we conducted simulations (described in Methods) to assess the accuracy of the displacement retrieval at low SBR for both methods. The chosen SBR corresponds to previous experimental SBR shown on Fig.3d. For the highest SBR, typically larger than 1.5, both methods are giving similarly good results, as the average displacement error is null in both cases (Fig.4a). However, as the SBR decreases, the centroid method is more impacted. A large dispersion of the shift measurement was observed in the case of the centroid method, as shown by the large error bars (Fig.4a) and Standard Error of the Mean (Fig.4b). Thus, the number of images over which the displacement calculation is done has to be increased by a factor 25 at a SBR of 1.2 to achieve the same precision on the displacement measurement with the centroid approach as with the correlation method. This variability in the displacement measurement affects directly the quality of the aberration correction that can be achieved with an AO loop. At SBR smaller than 1.2, the noise also impacts the estimate of the average displacement obtained through the centroid process (at SBR = 1.09, Student t-test $\alpha = 0.05$, p = 0.009). On the contrary, the mean value of the shift evaluated by the correlation-based measurement is not significantly different from the one imposed numerically at all SBR values assessed (at SBR = 1.09, Student t-test  $\alpha = 0.05$, p = 0.355). As the results provided in Fig.4a,b were obtained for a specific threshold of 0.1 for the centroid-based measurement (see Methods), we checked if they depend on the threshold value. For this purpose, the same simulation is computed for thresholds varying between 0.07 and 0.2 and the standard deviation and the mean values of the retrieved displacement error are represented respectively on Fig.4c and Fig.4d, as a function of the SBR. It shows that the threshold value of 0.1 was roughly optimal and that independently of the threshold on the centroid measurement, the shift retrieval performances of the correlation remain better, in low SBR conditions, whatever the threshold used in the centroid measurement. Our simulation proves thus that a much more accurate measurement of the shifts between thumbnails induced by aberration could be obtained at lower SBR with the correlation method rather than with the centroid-based method.
We also tested the accuracy of subpixelic shift retrieval on experimentally acquired thumbnails to take into account real scattering conditions from biological samples (see Methods). Results are reported on Fig.4a. Identically good shift retrieval is obtained with the two methods for SBR larger than 2; however, when the SBR drops, the error on the measured displacement remains within ± 0.1 pixel with the correlation method, whereas the centroid-based measurements deviates from zero, and this error increases significantly at SBRs smaller than 1.3. These results based on experimental data confirm the robustness of the shift retrieval for low SBR using the correlation method. 

\begin{figure}[ht]
    \centering
    \includegraphics[width=13cm]{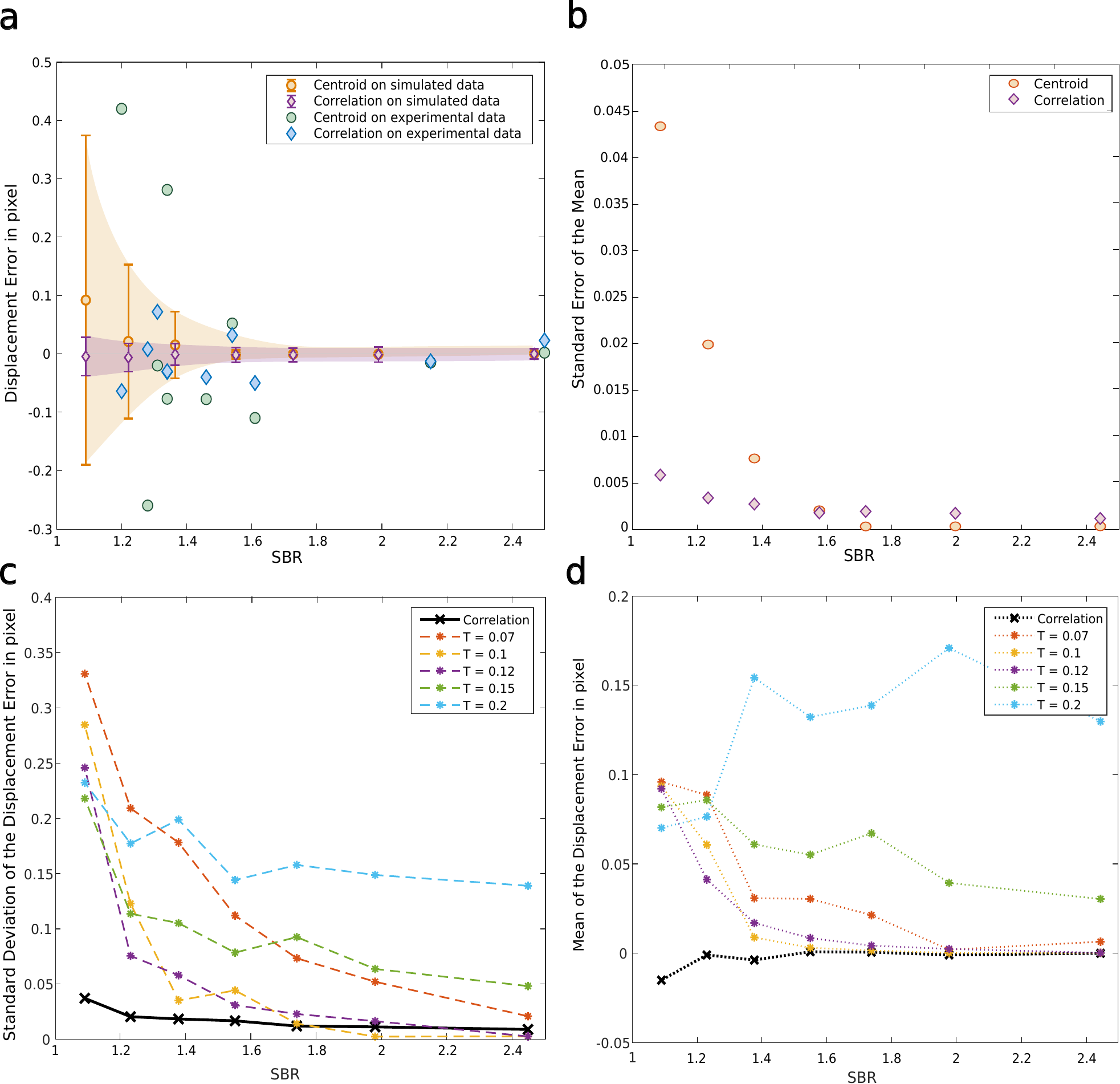}
    \caption{Comparison of shift retrieval accuracy using centroid and correlation method. (a) Numerical simulation of cross-correlation and centroid method to retrieve a subpixelic displacement induced numerically as a function of the SBR. Coloured dots with error bars correspond respectively to the mean and standard deviation for 50 simulations at each SBR. Single dots and diamonds represent shift retrieval for experimental images with both methods (b) Standard Error of the Mean (SEM) as a function of the SBR. The SEM values correspond to the simulated data in (a). (c-d) Standard deviation (c) and mean values (d) of the displacement error for 50 simulations at each SBR, for several thresholds (T) applied for the centroid method, T varying between 0.07 and 0.2 }
    \label{fig:Fig4}
\end{figure}

\subsection{ESSH allows closed-loop aberration correction for enhanced imaging through scattering media}
We finally tested the possibility to correct aberration and improve image quality in a closed-loop AO set up using the ESSH sensor (see Methods), with a model sample that resembles \emph{in vivo} biological samples in term of structures and photon flux. The sample consists in fluorescently-labelled neurons placed under a fixed unlabelled mouse brain slice (Fig.5a and Fig.5h, see Methods). Aberrations are induced by the brain slice and by the coverslip placed on top of it. The latter is generating a 210 nm SA3, prevailing to the 10 to 20 nm SA3 generated by the brain slice, the brain tissue mainly enabling the introduction of significant scattering. The geometry of the sample was chosen to mimic in depth optical sectioning in a scattering biological medium. The AO-loop improves significantly the image quality. The initial images (Fig.5b and Fig.5i) are corrected for the microscope aberrations, so that only correction of the sample-induced aberrations is performed to retrieve the images in Fig.5c and Fig.5j. For the thinnest sample, the signal from the soma is increased by 28\%, between Fig.5b and Fig.5c and the signal from dendrites is increased by 34\% (Fig.5f and Fig.5g). This contrast and signal enhancement reveals some dendrites that were not visible before the correction (Fig.5c). In the case of the thicker sample (thickness of $2.5 l_{s}$), the correction, which is achieved with a very low amount of ballistic photons (8\% of the incident light), increases the signal and contrast of the image even in these conditions (Fig.5j). In this proof of principle experiment, the gain in image quality is even significantly lower than the one expected using two-photon microscopy, as in the latter scattering on the emission path doesn’t affect image quality. This experiment illustrates the fact that a correction based on the ESSH wavefront measurement obtained through a scattering tissue and using fluorescently-labelled biological samples allows a significant image improvement, that is crucial especially in functional imaging where sensitivity is a key parameter.

\begin{figure}[ht]
    \centering
    \includegraphics[width = 13cm]{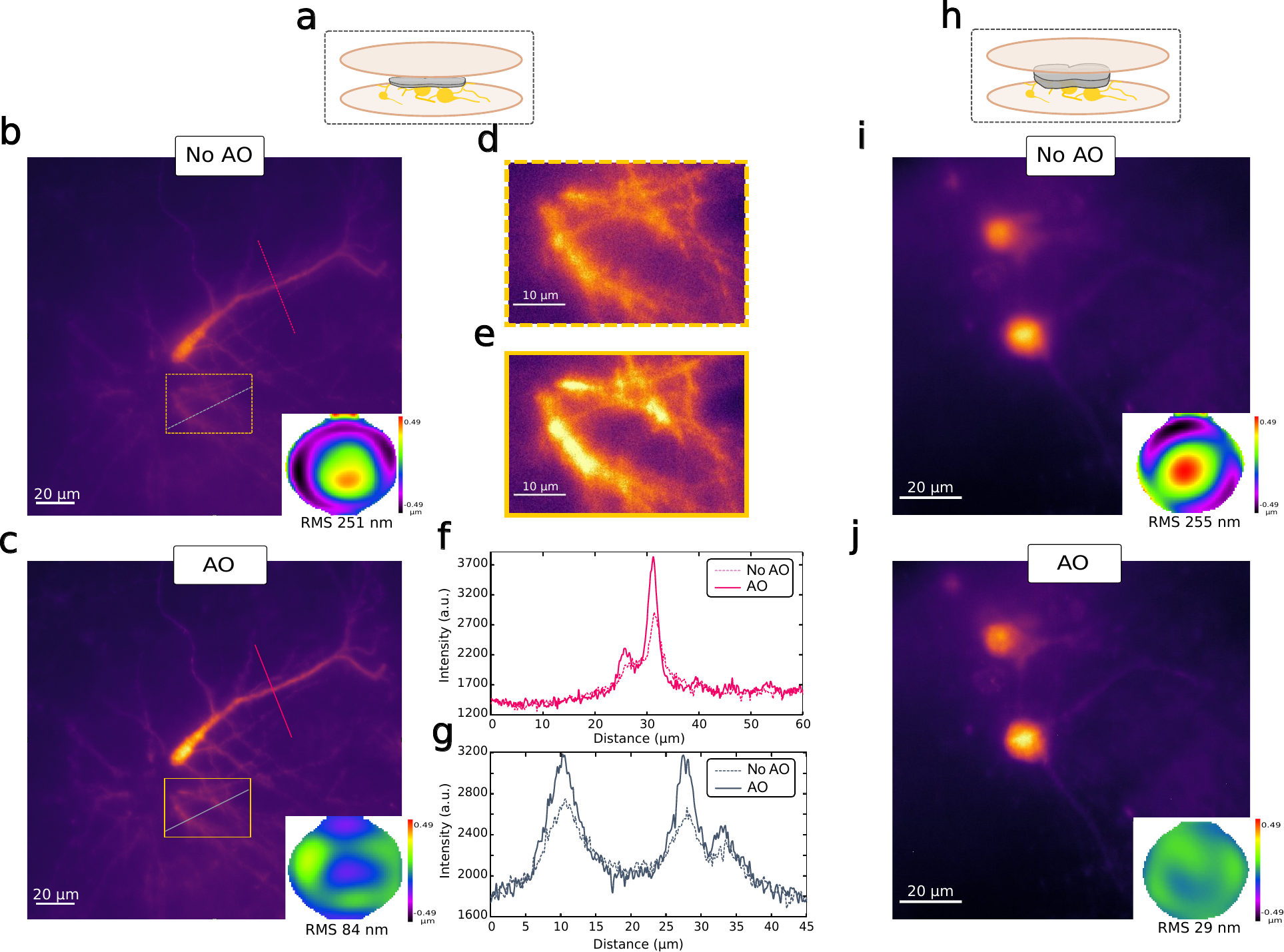}
    \caption{ESSH based AO correction loop enables to retrieve microscope performances through scattering samples. (a) Fixed mVenus neuron culture placed under a 50-µm fixed mouse brain slice, corresponding to $1.2 l_{s}$. Wide-field images before (b) and after the correction (c) with the corresponding wavefronts. Signal profiles along the black and pink lines in (b) and (c) are plotted in (f) and (g). Zoom on dendrites before (d) and after (e) correction, with adjusted contrast. (h) Same as (a) but using a 100-µm fixed mouse brain slice, corresponding to $2.5 l_{s}$. Images before (i) and after (j) correction, with corresponding wavefronts.}
    \label{fig:Fig5}
\end{figure}

\section{Discussion}\label{Discussion}
We showed that our innovative Shack-Hartmann sensor, based on an extended-source wavefront measurement, is a reliable tool to measure quantitatively aberrations in scattering samples up to $3.9   l_{s}$-thick. We have shown that wavefront measurements using Shack-Hartmann sensors is ultimately limited by the SBR on the thumbnails at the edge of the sensor, due to the increased scattering of the rays reaching them. However, by computing simulations at low SBR, corresponding to real scattering conditions, we obtained more accurate displacement measurements with the correlation-based method than with the centroid-based approach. This study demonstrates the robustness of the ESSH wavefront measurement approach in depth in biological samples with significant levels of scattering. This can be attributed to the use of correlation instead of centroid to estimate shifts. Concerning correlation, we currently make use of conventional correlation algorithms (see Methods), whereas more advanced correlation algorithms such as phase correlation are known to provide better sensitivity and/or accuracy, depending on the characteristics of the signal, but at the cost of more computational complexity and time. There is as a consequence still some room for further improvement of the approach. 

By digging into the characteristics of the ESSH signal, we have also shown that the use of the ESSH allows the measurement of the scattering lengths of these samples with a single image acquisition. Indeed, it allows to measure in a single shot using a single sample the attenuation through 7 effective thicknesses, corresponding to the 7 microlenses conjugated along the objective back focal plane radius. It facilitates therefore the measurement of $l_{s}$  as it can be obtained directly from a single measurement in one sample. Indeed, it doesn’t require multiple measurements at multiple depths\cite{Kobat_Durst_Nishimura_Wong_Schaffer_Xu_2009,Wang_Wu_Sinefeld_Li_Xia_Xu_2018} or the use of multiple samples of different thicknesses, which would introduce additional uncertainties associated with the estimation of each sample thickness. This single shot measurement of $l_{s}$  is therefore both accurate and easy to implement on a wide range of samples. A specific design of the ESSH with dedicated number of microlenses and characteristics of the camera could be realized to optimize the performance of the $l_{s}$ measurement. The accuracy of the measurement will depend on the sample thickness, as the relative attenuation between the central and border thumbnails in our case decreases as $\exp{(-0.5z_c/l_s)}-1$. Therefore, increasing the sample thickness is beneficial, until attenuation of the ballistic light is too strong to be separated from the background on each thumbnail. In our hand thicknesses of the order of $l_{s}$ to $3 l_{s}$ in fixed tissues seemed optimal.
The values obtained can be compared with the ones reported in the literature (see Table 1). Since scattering lengths measured \textit{in vivo} are 2 to 3 times higher than the one realized on fixed tissues \cite{Kobat_Durst_Nishimura_Wong_Schaffer_Xu_2009, Wang_Wu_Sinefeld_Li_Xia_Xu_2018}, we used for comparison values reported only in fixed slices\cite{Kobat_Durst_Nishimura_Wong_Schaffer_Xu_2009} and extrapolated linearly their measurement to the wavelengths used in our study. Our measurements match well the values reported in literature and confirm the ability of the ESSH sensor to provide a consistent measurement of $l_{s}$. The slight discrepancy can be related to differences in slice preparation or animal age. 

\begin{table}
    \centering
    \begin{tabular}{c|c|c|c}
         Sample & $\lambda$ [nm]  & $l_{s}$ [µm] & Reference \\
         \hline
         Fixed brain &  775 & 55.2 & Kobat et al.
         \cite{Kobat_Durst_Nishimura_Wong_Schaffer_Xu_2009}\\
         & 1280 & 106.4& \\
         \hline
         Fixed slice & 515 & 38 $\pm$ 2 & Results \\
         & 615 & 46 $\pm$ 2 & \\
         & 805 & 77 $\pm$ 5 &
         
    \end{tabular}
    \caption{Measured scattering lengths $l_{s}$ of fixed brain slices at different wavelengths with the ESSH analyzer compared to literature reported values.}
    \label{tab:Tab1}
\end{table}

One possible limitation of the scattering measurement method is the possible impact of the geometry of the sample on the measurement, in particular its surface. A non-flat surface will lead to some measurement artifacts, different local heights leading to differences in measured local scattering values. A proper use of the technique is then conditioned to the use of samples with flattened surfaces, for examples using glass slides or coverslips, which corresponds to a significant part of experimental conditions, such as e.g. transcranial imaging in the rodent using optical windows, which minimizes movement artifacts.

In this study we realized quantitative aberration measurement and correction in scattering conditions with an extended-source wavefront sensor, paving the way to the combination of adaptive optics based on this method to optical sectioning microscopy, such as two-photon microscopy or light-sheet microscopy, for deep imaging. We have shown here that the ESSH allows to perform quantitative aberration measurement deep inside biological tissues up to $3.9 l_{s}$-thick. Using standard GFP labelling ($\lambda_{em}$  = 514 nm), and since $l_{s}$ is of the order of 60 µm \emph{in vivo} \cite{Wang_Wu_Sinefeld_Li_Xia_Xu_2018}, the ESSH sensor would give access to quantitative aberration measurements, and then correction, up to $3.9 \times 60$µm$\approx 250$ µm deep inside a live mouse brain at this wavelength. To make ESSH-based wavefront measurement compatible with the maximum imaging depths in 2-photon microscopy (up to 800 µm in the mouse brain\cite{Liu_Li_Marvin_Kleinfeld_2019,Wang_Sun_Richie_Harvey_Betzig_Ji_2015}) or in light-sheet microscopy, a dual-labelling approach is relevant. It allows to keep the commonly used GFP-based functional labelling in the visible range, and to shift the wavelength used for the wavefront measurement towards the near-IR. This strategy presents two advantages: first, the impact of scattering is minimized on the ESSH thumbnails; secondly, the photon budget is optimized as all photons from functional dyes are kept for the neuronal activity recording. Using dye emitting near 900 µm, where $l_{s} \approx 170$ µm, AO corrections at depths of $3.9 \times 170$µm$ \approx 700$ µm are reachable. 

There are however several limitations to the present study that will require further experimental confirmation. First of all, the experimental demonstration realized in this study only partially confirms the prediction based on simulation. Indeed, the approach used to prepare the sample did not allow a precise enough control of the SBR of the wavefront measurement to correspond to extreme values (about 1.7 in our experiment), enabling only partial confirmation of the capabilities of the ESSH measurement to work with very low SBR. Also, the sample used does not reflect the behavior of signal and background for both light-sheet and 2-photon microscopy, since for example the main effect of aberrations in non-linear microscopy is a decrease of signal. Considering this, the proof of principle experiment is more representative of the behavior of the AO loop in a light-sheet experiment. This particular aspect will have to be investigated using a dedicated 2-photon setup. The present experiment is nevertheless a clear demonstration of the capability of an ESSH-driven AO loop to provide a correction in low SBR situations, whatever the origin of such SBR is, and on real biological objects of interest such as neurons. Also, there was no consideration here of the impact of the isoplanetic patch with the use of SA3, since this topic was already discussed in detail in our previous work\cite{Hubert_Harms_Juvenal_Treimany_Levecq_Loriette_Farkouh_Rouyer_Fragola_2019}.
Finally, our ESSH sensor doesn’t require descanning of the fluorescence, as in centroid measurement\cite{Wang_Sun_Richie_Harvey_Betzig_Ji_2015}, nor requires the use of a specific scanning arrangement and of an ultrafast laser for the wavefront measurement when the technique is applied outside non-linear microscopy \cite{Tao_Azucena_Fu_Zuo_Chen_Kubby_2011}. This is advantageous in term of photon budget, as the ESSH can be placed closer to the microscope objective. Furthermore, it makes it perfectly suited for advanced scanning methods as acousto-optic deflectors, where descanning is difficult to implement\cite{Akemann_Wolf_Villette_Mathieu_Tangara_Fodor_Ventalon_Leger_Dieudonne_Bourdieu_2022}.

\begin{backmatter}

\bmsection{Acknowledgments}
We thank Marine Depp, Ombeline Hoa, Alexandra Bogicevic, Rémi Carminati and Thomas Pons for technical assistance and fruitful discussion. This work was supported by grants from France’s Agence Nationale de la Recherche (INOVAO, ANR-18-CE19-0002). This work has received also support under the program « Investissements d’Avenir » launched by the French Government and implemented by ANR with the references ANR–10–LABX–54 (MEMOLIFE), ANR–10–IDEX–0001–02 (Université PSL) and ANR-10-INSB-04- 01 (France-BioImaging Infrastructure).

\bmsection{Disclosures}
F.H. is employed by the company Imagine Optic and A.H.’s doctoral research is funded by Imagine Optic. The other authors declare no competing interests.

\medskip

\bmsection{Data availability} Data underlying the results presented in this paper are not publicly available at this time but may be obtained from the authors upon reasonable request.

\end{backmatter}


\bibliography{sample}

\end{document}